\documentclass{aa}  

\usepackage{natbib}
\usepackage{graphicx}
\usepackage{txfonts}
\usepackage{hyperref}

\usepackage{amsmath}
\usepackage{nccmath}
\usepackage{subfig}
\usepackage{fixltx2e}
\usepackage[utf8]{inputenc}

\usepackage{xcolor}

\def\msun{\ifmmode M_{\odot} \else M$_{\odot}$\fi}
\def\msunyr{\ifmmode M_{\odot} {\rm yr}^{-1} \else M$_{\odot}$ yr$^{-1}$\fi}
\def\zsun{\ifmmode Z_{\odot} \else Z$_{\odot}$\fi}
\def\lsun{\ifmmode L_{\odot} \else L$_{\odot}$\fi}

\newcommand{\mstar}{\ifmmode M_\star \else $M_\star$\fi}
\newcommand{\luv}{\ifmmode L_{\rm UV} \else $L_{\rm UV}$\fi}
\newcommand{\lir}{\ifmmode L_{\rm IR} \else $L_{\rm IR}$\fi}
\newcommand{\lbol}{\ifmmode L_{\rm bol} \else $L_{\rm bol}$\fi}

\begin{document} 
  \title{Physics of a clumpy lensed galaxy at z=1.6}
  \author{M. Girard\inst{1}
     \and
           M. Dessauges-Zavadsky\inst{1}
     \and
           D. Schaerer\inst{1,2}
     \and
           J. Richard\inst{3} 
     \and
           K. Nakajima\inst{4,5} 
     \and
           A. Cava\inst{1}
          }

  \institute{Observatoire de Genève, Université de Genève,
              51 Ch. des Maillettes, 1290 Sauverny, Switzerland\\
              \email{marianne.girard@unige.ch}
         \and
             CNRS, IRAP, 14 Avenue E. Belin, 31400 Toulouse, France
         \and
           Univ Lyon, Univ Lyon1, Ens de Lyon, CNRS, Centre de Recherche Astrophysique de Lyon UMR5574, F-69230, Saint-Genis-Laval, France
          \and
             European Southern Observatory, Karl-Schwarzschildstasse 2, 85748 Garching, Germany
         \and
             National Astronomical Observatory of Japan, 2-21-1 Osawa, Mitaka, Tokyo 181-8588, Japan
             }

  \date{Received - ; accepted - }

  \abstract{Observations have shown that massive star-forming clumps are present in the internal structure of high-redshift galaxies. One way to study these clumps in detail with a higher spatial resolution is by exploiting the power of strong gravitational lensing which stretches images on the sky. In this work, we present an analysis of the clumpy galaxy A68-HLS115 at $z=1.5858$, located behind the cluster Abell 68, but strongly lensed by a cluster galaxy member. Resolved observations with SINFONI/VLT in the near-infrared (NIR) show  H$\alpha$, H$\beta$, [NII], and [OIII] emission lines. Combined with images covering the B band to the far-infrared (FIR) and CO(2-1) observations, this makes this galaxy one of the only sources for which such multi-band observations are available and for which it is possible to study the properties of resolved star-forming clumps and to perform a detailed analysis of the integrated properties, kinematics, and metallicity.
  We obtain a stability of $\upsilon_{rot}/\sigma_0 = 2.73$ by modeling the kinematics, which means that the galaxy is dominated by rotation, but this ratio also indicates that the disk is marginally stable. We find a high intrinsic velocity dispersion of $80\pm10$ km s$^{-1}$ that could be explained by the high gas fraction of $f_{gas}=0.75\pm0.15$ observed in this galaxy. This high $f_{gas}$ and the observed sSFR of $\rm 3.12 \, Gyr^{-1}$ suggest that the disk turbulence and instabilities are mostly regulated by incoming gas (available gas reservoir for star formation). The direct measure of the Toomre stability criterion of $Q_{crit}=0.70$ could also indicate the presence of a quasi-stable thick disk. Finally, we identify three clumps in the H$\alpha$ map which have similar velocity dispersions, metallicities, and seem to be embedded in the rotating disk. These three clumps contribute together to $\sim40\%$ on the SFR$_{H\alpha}$ of the galaxy and show a star formation rate density about $\sim100$ times higher than HII regions in the local Universe. 
  }

  \keywords{galaxies: high redshift -- galaxies: kinematics and dynamics}

\maketitle


\section{Introduction}

Clumpy galaxies have been observed for the first time between redshift 0.5 and 3 by \citet{Cowie1995} using Hubble Space Telescope (HST) high-resolution imaging. 
More recent observations using deep imaging, for example in the Hubble Ultra Deep Field, have confirmed the presence of these star-forming clumps in the internal structure of the galaxy disks \citep[e.g.][]{Elmegreen2007, Elmegreen2009, Livermore2012}. 
Integral field unit (IFU) observations have also played an important role in our understanding of these clumps and their host galaxies by providing key information about their kinematics and physical properties \citep[e.g.][]{ForsterSchreiber2009, Forster-Schreiber2011, Genzel2011, Wisnioski2012, Livermore2015}.

The size of these clumps has been reported to range from a few dozen parsecs to a few kiloparsecs from  observations at $z=1-2$ \citep[e.g.][]{Livermore2012,Swinbank2012,Livermore2015,Cava2018},
 whereas the mass lies around $10^{8-9} \mathrm{M}_{\odot}$ \citep[e.g.][]{Forster-Schreiber2011, Guo2012}.
However, it has been shown recently by \citet{Dessauges-Zavadsky2017a} that the masses of these clumps are probably significantly lower, by  one or more orders of magnitude ($\sim10^7$ \msun). 
Indeed, a limited spatial resolution could cause an increase of the mass obtained due to the clustering of clumps of smaller masses, and, more importantly, the sensitivity limit impacts the clump masses even more strongly by inducing a bias against the low-mass clumps \citep{Dessauges-Zavadsky2017a, Tamburello2017, Cava2018}.

The properties of these massive clumps differ from what we observe in local HII regions. Using H$\alpha$ observations, \citet{Jones2010} find that the star formation rate (SFR) density can reach 100 times higher values for clumps at $z=1.7-3.1$. From CO observations, \citet{Canameras2017} find SFR densities up to 2000 $\rm M_\odot \, yr^{-1} \, kpc^{-2}$ for clumps of a massive galaxy at $z=1.5$, which is in the range of maximal starbursts. 
Their formation is also probably due to internal processes in the galaxy disk instead of mergers. Merging systems rather than internal gravitational instabilities would result in disturbing the dynamics of the disk, while the observations and simulations often show clumps which are embedded in a rotating disk \citep[e.g.][]{Forster-Schreiber2011, Bournaud2014, Tamburello2015, Livermore2015, Mandelker2017, Cava2018}. It is believed that these clumps are formed through fragmentation processes of unstable disks \citep{Dekel2009}.

   
      \begin{figure*}[t]
   \centering
   \subfloat{\includegraphics[scale=0.52]{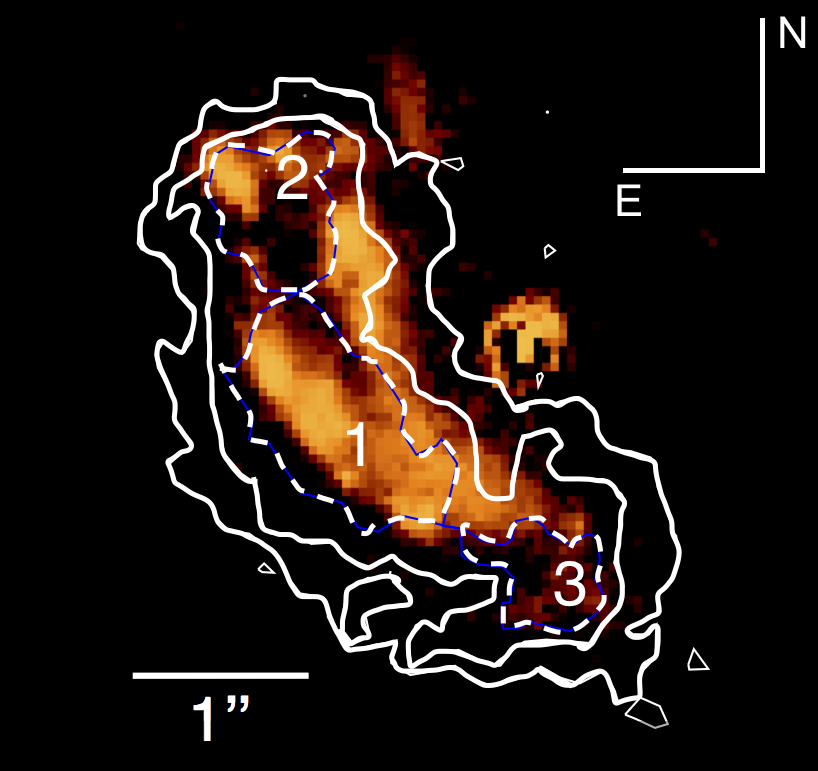}}
   \subfloat{\includegraphics[scale=0.74]{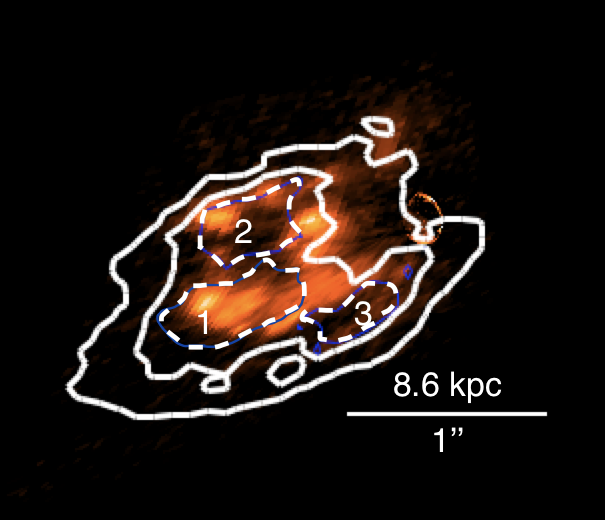}}
      \caption{HST/F814W images of A68-HLS115 in the image (left panel) and source plane after reconstruction (right panel). White dashed contours represent the H$\alpha$ clumps and the white solid contours the H$\alpha$ surface brightness measured in the image (left panel) and source plane (right panel) surface brightness maps corresponding to 0.64 and 1.28 $\rm \times 10^{-16} \, erg \, s^{-1} \, cm^{-2} \, arcsec^{-2} $. 
      The surface brightness is conserved between the image and source plane.
     The cluster galaxy which is acting as a lens has been removed from the two images.}
         \label{HST_image}
   \end{figure*}

To better understand the nature of these clumps and the role they play in the galaxy evolution, several studies have established different scaling relations using the clump properties.
It has been found that their surface brightness is evolving with redshift, which could be explained by the higher gas fraction observed in high-redshift galaxies and the fragmentation on larger scales in these systems (\citeauthor{Livermore2012} \citeyear{Livermore2012}, \citeyear{Livermore2015}). By comparing the clump properties to the properties of the host galaxies, \citet{Livermore2012} find a scaling relation between the clump surface brightness and the surface density of the host galaxy, meaning that the properties of these clumps are strongly related to the host galaxy. Also, \citet{Dessauges-Zavadsky2017a} have shown that the maximum clump mass is correlated to the host mass.

Nevertheless, smaller clumps are often not resolved by direct observations and it is then impossible to study their physical properties. One way to study these clumps in detail with a better spatial resolution is by using strong gravitational lensing which stretches images on the sky  \citep[e.g.][]{Jones2010, Swinbank2011, Livermore2012, Wuyts2012, Adamo2013, Livermore2015,  Dessauges-Zavadsky2017a, Cava2018, Patricio2018}. In addition to a better spatial resolution, the magnification effect caused by the lens allows to one obtain a better signal-to-noise ratio (S/N) than direct observations for a galaxy with the same luminosity.

Gravitational lensing can also provide more information about the kinematics and metallicity of galaxies and can allow to reach a lower stellar mass range of galaxies more difficult to observe with direct observations \citep[e.g.][]{Jones2010, Livermore2015, Leethochawalit2016, Mason2017,Girard2018, Patricio2018}. \citet{Leethochawalit2016} point out that a high spatial resolution can reveal much more complex kinematics and metallicity gradient. Their work, which uses strong gravitational lensing and studies low-mass galaxies ($\rm log(M_\star/M_\odot)\sim9.5$) at $z\sim2$, does not agree with a simple rotation disk model most of the time. They obtain a fraction of only 36\% rotation-dominated galaxies in their sample.

In this work, we present an analysis of the galaxy A68-HLS115 at $z=1.5858$, located behind the galaxy cluster Abell 68, but strongly lensed by a cluster galaxy member. A detailed study of the integrated physical properties of this galaxy has already been performed, exploiting the detection of this galaxy through the B band up to the far-infrared (FIR) band\footnote{B band from CFHT/12k, F702W and F814W from HST, z band from FORS3/VLT, J and H bands from ISAAC, Ks photometry from UKIRT, 3.6 $\mu$m and 4.5 $\mu$m from Spitzer/IRAC, 24 $\mu$m from MIPS, 100 $\mu$m and 160 $\mu$m from Herschel/PACS, and 250 $\mu$m, 350 $\mu$m and 500 $\mu$m from Herschel/SPIRE} \citep{Sklias2014} and in CO(2-1) observations with the IRAM interferometer at the Plateau de Bure, France \citep{Dessauges-Zavadsky2015}.

By combining our IFU observations obtained with SINFONI/VLT to these previous studies, this makes this galaxy one of the only sources for which such multi-band observations are available and for which it is possible to study the properties of resolved star-forming clumps and to perform at the same time a detailed analysis of the galaxy metallicity and kinematics. This source is also one of the most gas-rich galaxies (with a molecular gas fraction of 75\%) known at redshifts higher than one.

The paper is organized as follows. Section \ref{Sect2} describes the results of previous studies of the galaxy A68-HLS115 which have been done by our group (\citeauthor{Sklias2014} \citeyear {Sklias2014}; \citeauthor{Dessauges-Zavadsky2015} \citeyear{Dessauges-Zavadsky2015}). In Sect. \ref{Sect3}, we present SINFONI/VLT observations and data reduction. In Sect. \ref{Sect4}, we explain the lens modeling and measurements of the emission lines. Section \ref{Sect5} presents our result on the integrated physical properties, metallicity, kinematics, and properties of clumps of the galaxy. We finally present our conclusions in Sect. \ref{Sect6}.

In this paper, we use a cosmology with : H$_0=70$ km s$^{-1}$ Mpc$^{-1}$, $\Omega_M=0.3$, and $\Omega_{\Lambda}=0.7$. When using values calculated with the initial mass function (IMF) of \citet{Salpeter1955}, we correct by a factor 1.7 to convert to a \citet{Chabrier2003} IMF.

   
      \begin{figure*}[htb]
      \centering
   \includegraphics[scale=1]{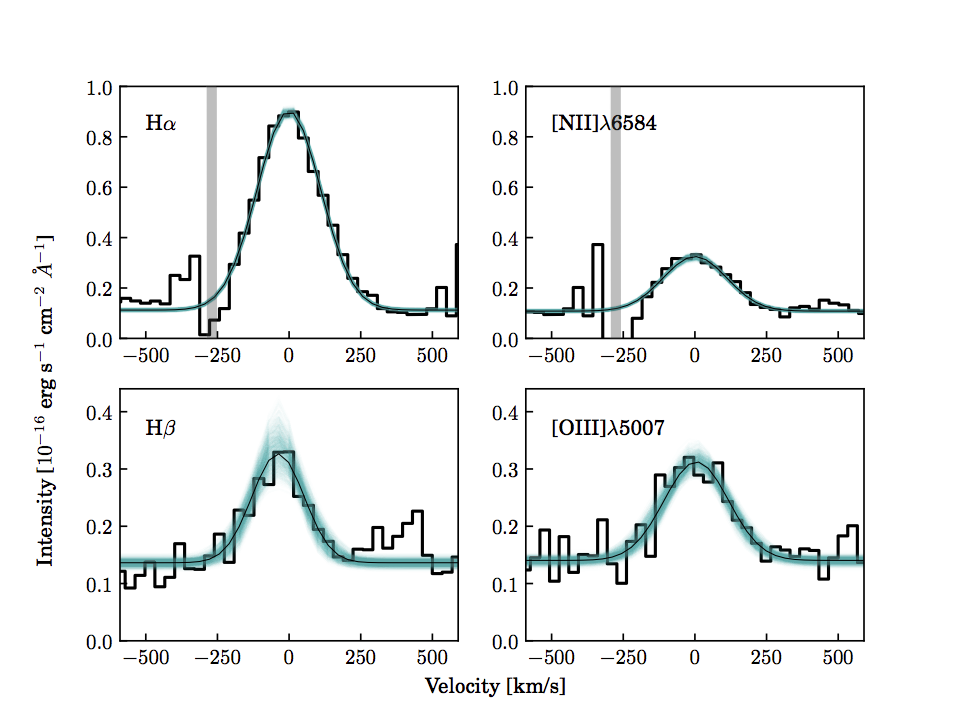}
    \caption{Integrated H$\alpha$, H$\beta$, [NII] and [OIII] emission lines from the SINFONI/VLT data in the image plane. The zero velocity corresponds to the center of the H$\alpha$ emission line. The 1000 realizations obtained with the Monte Carlo simulation are shown in blue and the best fit is shown in black. The gray lines show the positions of skylines.}
         \label{integrated_spectrum}
   \end{figure*}

   \begin{table}[htb]
     \caption[]{Physical properties of A68-HLS115 from \citet{Sklias2014} and \citet{Dessauges-Zavadsky2015}.}
         \label{CO}
\centering                          
\begin{tabular}{l c }       
\hline\hline                 
 Parameters     &  Values \\
\hline                        
            \noalign{\smallskip}
            $L_{\mathrm{IR}} \; [10^{12} \; \mathrm{L}_{\odot}] $    &    $ 1.12 \pm 0.06 $      \\
            $L_{\mathrm{UV}} \; [10^{10} \; \mathrm{L}_{\odot}] $    &    $ 2.17 \pm 0.20 $      \\
            M$_{\star} \; [10^{10}  \; \mathrm{M}_{\odot}] $   &    $ 2.63 ^{+0.54}_{-0.65} $       \\
             SFR$_{\mathrm{SED}} \; [\mathrm{M}_{\odot} \mathrm{yr}^{-1}] $  &    $ 81.4 ^{+8.5}_{-19} $          \\
            SFR(IR+UV)  $[\mathrm{M}_{\odot} \mathrm{yr}^{-1}]$            &  $118 \pm 6 $\\
            sSFR= SFR$_{\mathrm{SED}} / \mathrm{M}_{\star} \; [\mathrm{Gyr}^{-1}] $   & $3.12 $                   \\
            $A\rm_{V \, IR/UV}$\tablefootmark{a}               &    $ 1.58^{+0.30}_{-0.10}$                                                    \\
            T$_{\mathrm{dust}}$ [K]             &    $37.5 \pm 1    $                           \\
               $z_{CO}$                                  &   $1.5859 $               \\
            F$_{CO}$\tablefootmark{b} [Jy km s$^{-1}$]                     &   $2.00  \pm 0.30  $ \\
            M$_{\mathrm{gas}}$\tablefootmark{c} $[10^{10} \mathrm{ M}_{\odot}]$    &   7.83                   \\
            $f_{gas}$                        & $0.75 \pm 0.15  $             \\
            
            \noalign{\smallskip}
\hline    
\end{tabular}
  \tablefoot{The values have been adjusted to the Chabrier IMF and corrected for a lensing magnification of $\mu=4.6$ (see Sect. \ref{lens}).
\tablefoottext{a}{$A_{\mathrm{V}}$ is obtained from the ratio of $L_{\mathrm{IR}}$ over $L_{\mathrm{UV}}$ as discussed in \citet{Sklias2014} and \citet{Schaerer2013}.}
\tablefoottext{b}{Observed CO(2-1) line integrated flux.}
\tablefoottext{c}{M$_{\mathrm{gas}}$ is obtained assuming a CO-to-H$_2$ conversion factor of 4.36~$\rm {M}_{\odot}/K \, km\, s^{-1} \, pc^2$.}}
\end{table}


\section{The galaxy A68-HLS115}
\label{Sect2}

\citet{Sklias2014} have derived several physical properties of A68-HLS115 from the spectral energy distribution (SED), such as the infrared (IR) luminosity integrated from 8 $\mu$m to 1000 $\mu$m, the dust temperature, T$_\mathrm{dust}$, the ultraviolet (UV) luminosity, $L_{\mathrm{UV}}$, the stellar mass, $\rm M_\star $, the SFR, the specific star formation rate, sSFR,  and the extinction, ${A}_V $. We present these quantities in Table \ref{CO}. Overall, the analysis reveals a young galaxy of $\sim130$ Myr still actively producing new stars since a recent starburst took place. This galaxy of $\rm M_\star = 2.63^{+0.54}_{-0.65} \times 10^{10} \, M_\odot$ and $\rm SFR_{SED}=81.4^{+8.5}_{-19} \, M_\odot yr^{-1}$ also lies above the main sequence (MS), computed at the same redshift and stellar mass, with an offset of 0.3 dex, placing it within the accepted thickness of the MS, 0.3 < sSFR/sSFR$_{MS}$ <  3 (e.g., \citeauthor{Daddi2007} \citeyear{Daddi2007}; \citeauthor{Rodighiero2010} \citeyear{Rodighiero2010}; \citeauthor{Salmi2012} \citeyear{Salmi2012}). Figure \ref{HST_image} shows the HST/F814W images in the image and source plane after the reconstruction. 

The integrated CO emission properties of A68-HLS115, that is, redshift, observed CO(2-1) flux, molecular gas mass, $\rm M_{gas} $, and molecular gas fraction, $f_{gas}$, from \citet{Dessauges-Zavadsky2015} are also listed in Table \ref{CO}. The CO emission analysis shows a very high gas fraction ($f_{gas}=0.75\pm0.15$), making this galaxy one of the most gas-rich known for a redshift larger than one. 

In this work, we use a total magnification factor of $\mu= 4.6 \pm 0.4$. This value is different from the one used in \citet{Sklias2014} and \citet{Dessauges-Zavadsky2015} since we did further adjustments to improve the gravitational lens model (see Sect. \ref{lens}). All the physical parameters from \citet{Sklias2014} and \citet{Dessauges-Zavadsky2015} presented in this work have been corrected for this new magnification.

\section{Observations and data reduction}
\label{Sect3}

The near-infrared (NIR) IFU observations were undertaken with SINFONI on VLT
in service mode (ID: 092.B-0677(A), PI; Zamojski) on October Oct 6--8, 2013.
SINFONI was used in the seeing limited mode, and adopted the
$8^{\prime\prime} \times 8^{\prime\prime}$ field of view of the widest
$0.25^{\prime}$ scale. SINFONI IFU was taken using the J- and H-band
filters sampling the wavelength ranges of $1.1$--$1.4$ and
$1.45$--$1.85\,\mu$m with the resolving power of $R\sim 2000$ and
$\sim 3000$, respectively. Individual exposures of $300$\,s were
taken in both J- and H-band. We adopted two ABBA dithering patterns of eight
exposures per OB to perform sky subtraction with two positions, while
keeping the galaxy within the field-of-view of the instrument. The
seeing during the observations was $\sim
0.6^{\prime\prime}$--$0.8^{\prime\prime}$. Two OBs in J-band and two in H-band resulted 
in a total on-source integration time of $1.3$\,hours in
each band.

The SINFONI data was reduced using recipes in the standard ESO SINFONI
pipeline (v.2.5.2) operated through esoreflex (v.2.8). The pipeline
corrected for sky background, flat field, and distortions, and spectrally
calibrated each individual slice before reconstructing a 3D data
cube for each individual exposure data. We additionally performed a
sky background subtraction for an A-position image by subtracting an
average B-position image created from the previous and following
images, both of which were optimally scaled. This was feasible because
the dither size was significantly larger than our science source size.
We aligned individual cubes in the spatial direction and made a
composite by averaging them with a $2\sigma$-clipping rejection. Flux
solutions and telluric absorption correction were obtained from B8- and
B9-type bright stars ($K_{\rm Vega}=7.2$--$7.5$) observed at similar
times and airmasses. The intrinsic spectra of the stars were removed
by dividing the observed stellar spectrum by the templates created by
a stellar spectral synthesis program \citep[SPECTRUM;][]{Gray1994}
based on the \citet{Kurucz1993}'s atmosphere models. We finally registered
the processed J- and H-band cubes to the astrometry of the HST/F814W
image by using the several bright objects commonly detected in the
F814 image and SINFONI 2D images, which were created from the 3D cubes
collapsed in the wavelength direction.


\begin{table}[tb] 
\caption{Integrated and kinematic properties.}            
\label{int}      
\centering                          
\begin{tabular}{l c }        
\hline\hline                 
Parameters     &  Values \\
\hline                        
            \noalign{\smallskip}
            Redshift                &  $ 1.58582 \pm 0.00005  $ \\
            Total magnification     &  $ 4.6 \pm 0.4          $ \\
            SFR$_{H\alpha}\tablefootmark{a}  [\mathrm{M}_{\odot} \mathrm{yr}^{-1}] $  &  $ 202 \pm 51             $ \\
            $A\rm _{V \, \mathrm{H}\alpha/\mathrm{H}\beta}$              & $2.90 \pm 0.58 $           \\
            $A\rm _{V \, \mathrm{IR/UV}}$ /A$\rm _{V \, \mathrm{H}\alpha/\mathrm{H}\beta }$ & 0.54          \\
            12+log(O/H)$_{N2}$         & $ 8.58 \pm 0.07    $        \\
            12+log(O/H)$_{O3N2}$       & $ 8.52 \pm 0.10     $       \\
            $\rm \sigma_{int} \,  [km s^{-1}] $            & $ 109 \pm 3$       \\
            \noalign{\smallskip}
\hline                                   
            \noalign{\smallskip}
            $\rm \sigma_0 \, [km \, s^{-1}]$             &    $\rm 80.5 \pm 10$  \\ 
            $\rm \upsilon_{rot} \, [km \, s^{-1}]$       &    $\rm 218^{+12}_{-31}$  \\
            $\rm \upsilon_{rot}/\sigma_0$ &    $\rm 2.73$  \\
            $\rm PA\tablefootmark{b} \, [^{\circ}]$ &    $\rm -25\pm7$  \\
            $\rm i \, [^\circ]$ &  $48\pm15$  \\
            \noalign{\smallskip}

\hline                                   

\end{tabular}
 \tablefoot{
    \tablefoottext{a}{The SFR has been corrected for lensing and dust attenuation}.\\
     \tablefoottext{b}{The position angle, PA, is defined as $0^{\circ}$ for the north (up) and $90^{\circ}$ for the East (left).}
     }
\end{table}


\section{Analysis}
\label{Sect4}

\subsection{Gravitational lens modeling}
\label{lens}

We look into modelling the lensing effects of A68-HLS115 to reconstruct its intrinsic morphology. This galaxy is lensed by the combination of the cluster Abell 68 on a large scale, at a distance of 42 arcsec from the BCG, as well as the very near cluster member (residuals seen in Fig.\ref{HST_image}). We use the well-constrained mass model of Abell 68 which was presented in detail in \citet{Richard2007} and later improved in \citet{Richard2010}. Following the scaling relations found for cluster members, this model predicts a velocity dispersion of $\sim$150 km s$^{-1}$ for the cluster member affecting HLS115, based on its luminosity. However, for such a mass the emission peaks seen in continuum and H$\alpha$ images (Fig.\ref{HST_image}) would have symmetric pairs which are not detected in the images. We use the curvature of the emission in A68-HLS115 and the lack of multiple images to constrain this velocity dispersion to 80  km s$^{-1}$, as a lower value would make the reconstructed source highly elongated. Our model is consistent with the presence of a low-surface-brightness Einstein radius surrounding the lensing galaxy, as seen in the H$\alpha$ map in the image plane. Because of the uncertainty in the velocity dispersion, we derive a total magnification factor of $\mu=4.6\pm0.4$ for A68-HLS115 and the derived parameters (such as the physical scales) have a typical uncertainty of 10\%. 
This magnification factor is different from the one used in \citet{Sklias2014} and \citet{Dessauges-Zavadsky2015} of $\mu\sim15$ since they mostly used the original model from \citet{Richard2007} without any adjustment.

\subsection{Emission line measurement}

\label{Sect4.2}

The H$\alpha$, H$\beta$, [NII]$\lambda6584$, and [OIII]$\lambda5007$
emission lines are detected in the integrated spectrum and the main emission lines obtained in the image plane are presented in Fig. \ref{integrated_spectrum}.
Each emission line is fitted individually following a Gaussian model and using a $\chi^2$ minimization (Levenberg-Marquardt algorithm), except [NII]$\lambda6584$ for which we impose the same full-width at half maximum (FWHM) and relative velocity as the H$\alpha$ line since one side of the [NII]$\lambda6584$ emission line is strongly affected by a skyline. All pixels affected by the skylines are masked during the fitting process. To determine the uncertainties, we perform Monte Carlo simulations perturbing the flux of every pixel with an error defined as the standard deviation of the continuum around the emission line and fitting 1000 realizations per emission line as shown in blue in Fig. \ref{integrated_spectrum}. As a result, we obtain an accurate redshift, the flux, and the FWHM of each emission line and their associated uncertainties. We find observed fluxes of $12.45\pm0.22$, $1.83\pm0.23$,  $3.43\pm0.16$, and $\rm 2.10\pm0.32$ in units of $\rm 10^{-16} \, erg \, s^{-1} \, cm^{-2}$ for H$\alpha$, H$\beta$, [NII]$\lambda6584$, and [OIII]$\lambda5007$, respectively.

To determine the H$\alpha$ luminosity, we first correct the H$\alpha$ flux for the foreground dust attenuation caused by the Galaxy (using the value from the NASA/IPAC Extragalactic Database). 
We then correct for the dust attenuation in the galaxy itself with the Balmer decrement (H$\alpha$/H$\beta$) and the relation from \citet{Calzetti2000}. We use a $R_V=4.05$ 
and derive the star formation rate, SFR$_{H\alpha}$, following the \citet{Kennicutt1998} equation:

\begin{ceqn}
\begin{align}
\label{eq_sfr}
\mathrm{SFR}_{H\alpha} = 7.9 \times 10^{-42} \; {L(H}\alpha) \times \frac{1}{1.7} \times \frac{1}{\mu}, 
\end{align}
\end{ceqn}
where the factor 1.7 is the correction for the \citet{Chabrier2003} IMF, and  $\mu$ is the magnification. As a result, we obtain a lensing-corrected value of SFR$_{H\alpha}$= $\rm 202\pm51 \, M_\odot \, yr^{-1}$.
The main source of uncertainty of SFR$_{H\alpha}$ is due to the extinction correction, which is high (see Sect. \ref{Sect51}).

We derive the integrated velocity dispersion, $\sigma_{int}$, from the FWHM ($ \sigma_{obs}= c/\lambda_{obs} \times \mathrm{FWHM}/2.355$, where $c$ is the light speed and $\lambda_{obs}$ is the wavelength of the observed line) and by correcting the observed velocity dispersion, $\sigma_{obs}$, for the instrumental broadening, $\sigma_{instr}$, which is obtained from the skylines:

\begin{ceqn}
\begin{align}
\label{eq_sigma}
\sigma_{int} = \sqrt{\sigma_{obs}^2 - \sigma_{instr}^2}.
\end{align}
\end{ceqn}
We get a value of $\sigma_{int}= 109 \pm 3$  km s$^{-1}$ from the H$\alpha$ emission line. 
\\

\begin{figure*}
   \centering
   \subfloat{\includegraphics[scale=0.38]{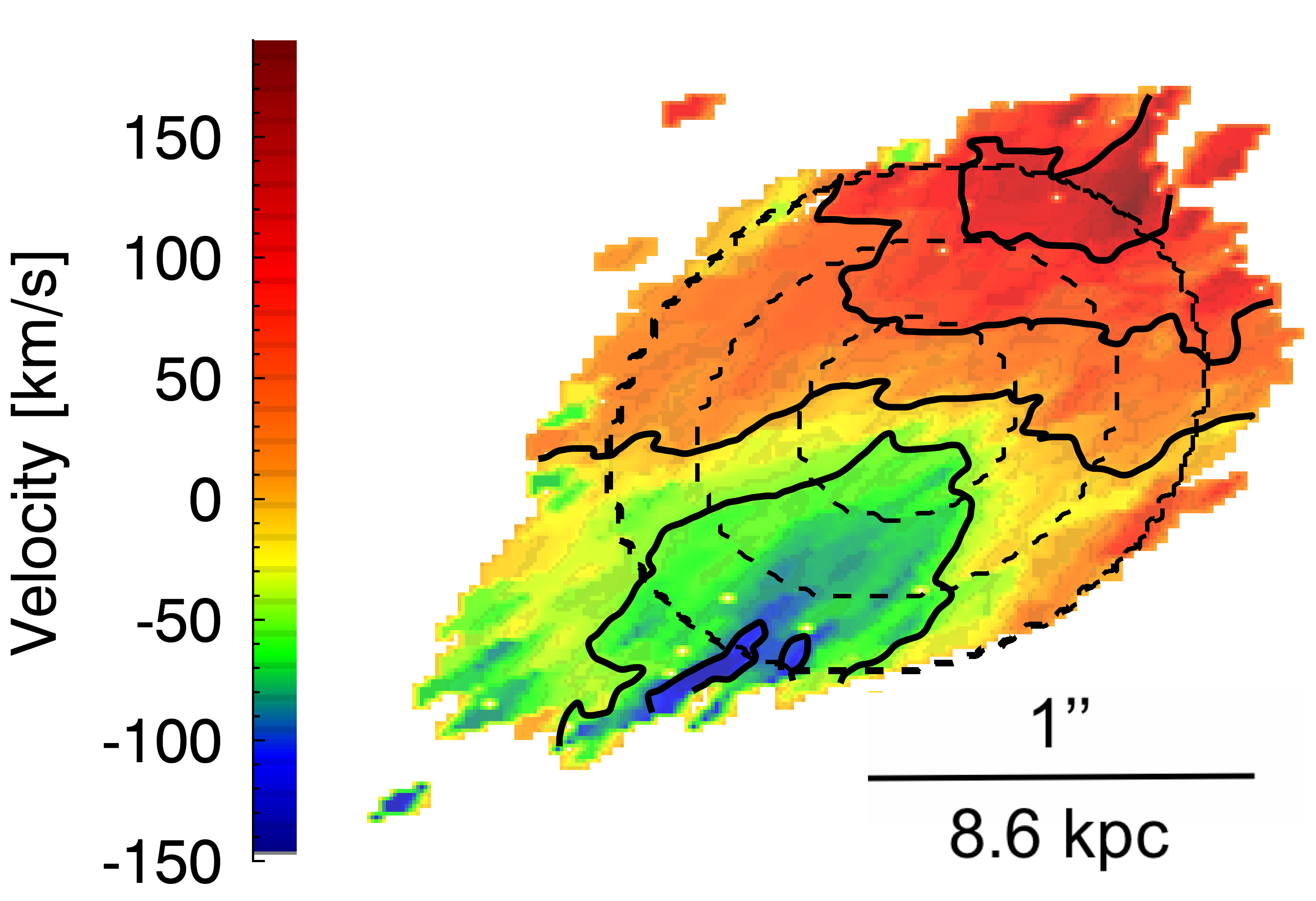}}
   \hspace{0.2cm}
   \subfloat{\includegraphics[scale=0.28]{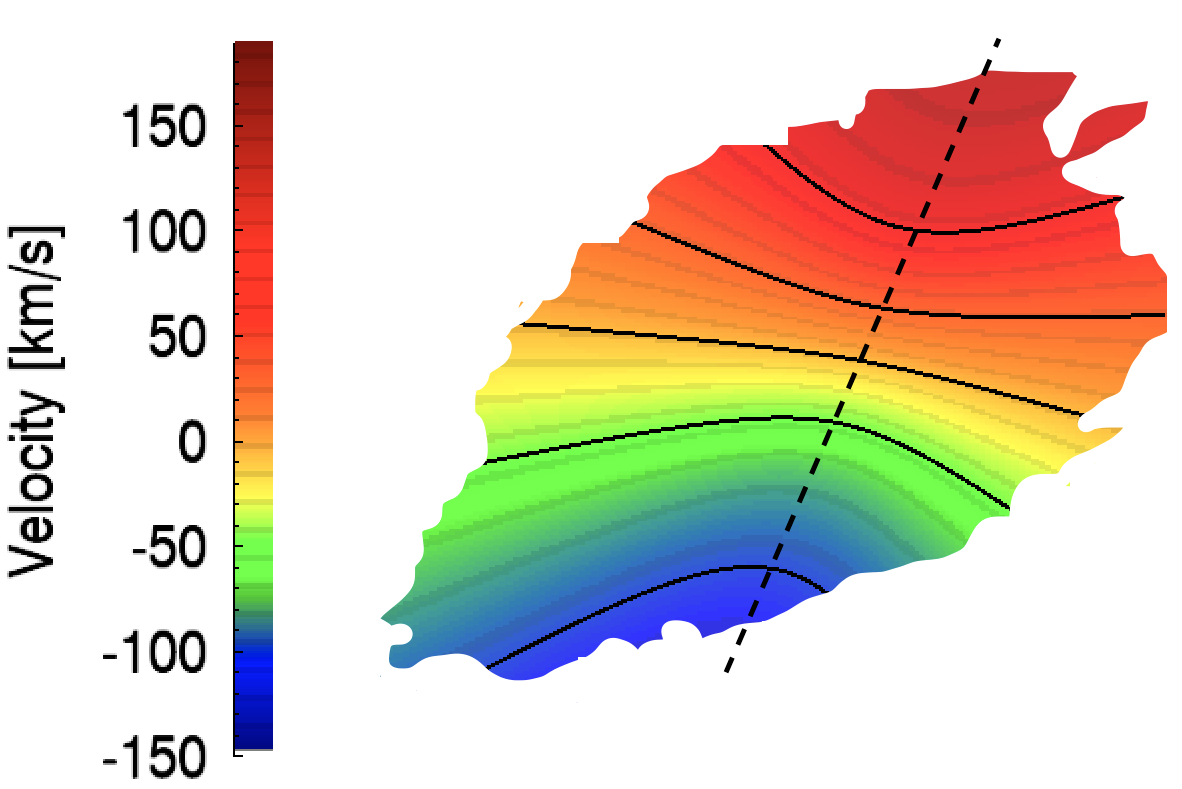}}
   \hspace{0.2cm} 
    \subfloat{\includegraphics[scale=0.215]{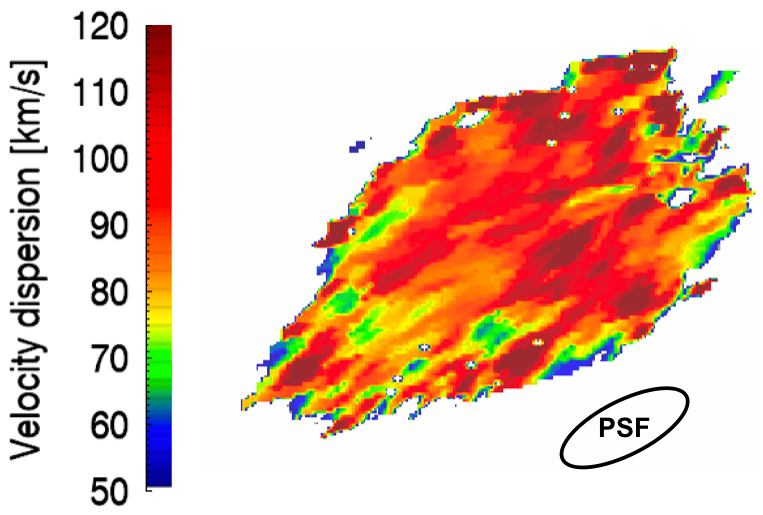}}
      \caption{Observed velocity (left panel), velocity from the model (middle panel), and velocity dispersion (right panel) in the source plane. Black lines in the velocity maps represent velocities of $-100$, $-50$, 0, 50 and 100  km s$^{-1}$. The black dashed ellipses (left panel) show the three annuli obtained using the \citet{Krajnovic2006} method to derive the metallicity gradient. The black dashed line (middle panel) represents the major axis.}
         \label{kin_source}
   \end{figure*}

We are also able to perform Gaussian fits, in the same way as the integrated spectrum, on the H$\alpha$ and [NII]$\lambda6584$ emission lines in individual spaxels in the image plane to obtain the flux, the [NII]/H$\alpha$ ratio, velocity, and velocity dispersion maps. We reject the spaxels where the S/N is lower than three. The H$\beta$ and [OIII]$\lambda5008$ emission lines are too faint in individual spaxels. We used the Lenstool best model to reconstruct the measured maps (H$\alpha$ flux, the [NII]/H$\alpha$ ratio, S/N, and kinematics) into a regular grid in the source plane, 
fixing the source plane spaxel grid to one fifth of the image plane SINFONI spaxels to follow the magnification factor along the shear direction (see Figs.\ref{flux_ha} and Fig.  \ref{kin_source}, respectively). This way we kept a similar sampling of the point spread function (PSF) to the image plane in the direction of the best resolution.

   
      \begin{figure}
   \centering
   \subfloat{\includegraphics[scale=0.94]{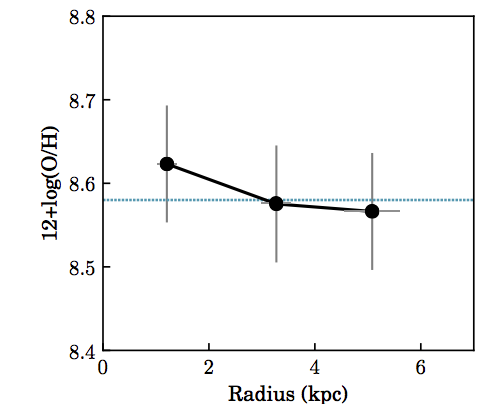}}\\
   \subfloat{\includegraphics[scale=0.94]{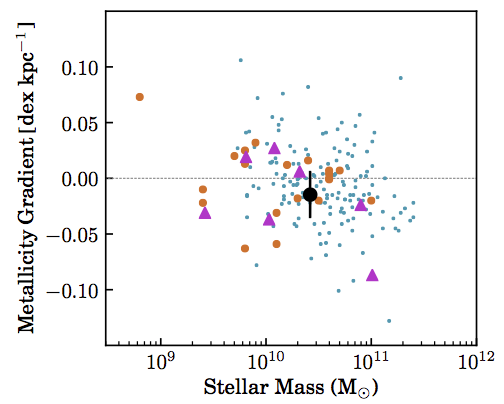}}\\ \subfloat{\includegraphics[scale=0.94]{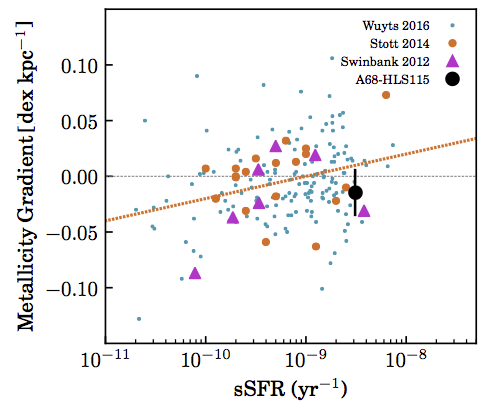}}
      \caption{Oxygen abundance as a function of the radius derived from \citet{Pettini2004} N2 indicator (top panel), metallicity gradient as a function of stellar mass (middle panel), and metallicity gradient as a function of sSFR (bottom panel). The blue line represents the oxygen abundance derived from the integrated spectrum. The gray lines indicate a value for the metallicity gradient of zero and the orange line is the relation obtained by \citet{Stott2014} from the combination of their data, \citet{Rupke2010b}, \citet{Swinbank2012}, and \citet{Queyrel2012}. We include the samples of \citet{Swinbank2012}, \citet{Stott2014}, and \citet{Wuyts2016} for comparison.
      }
         \label{met_gradient}
   \end{figure}

   
      \begin{figure}
   \centering
        \includegraphics[scale=0.3]{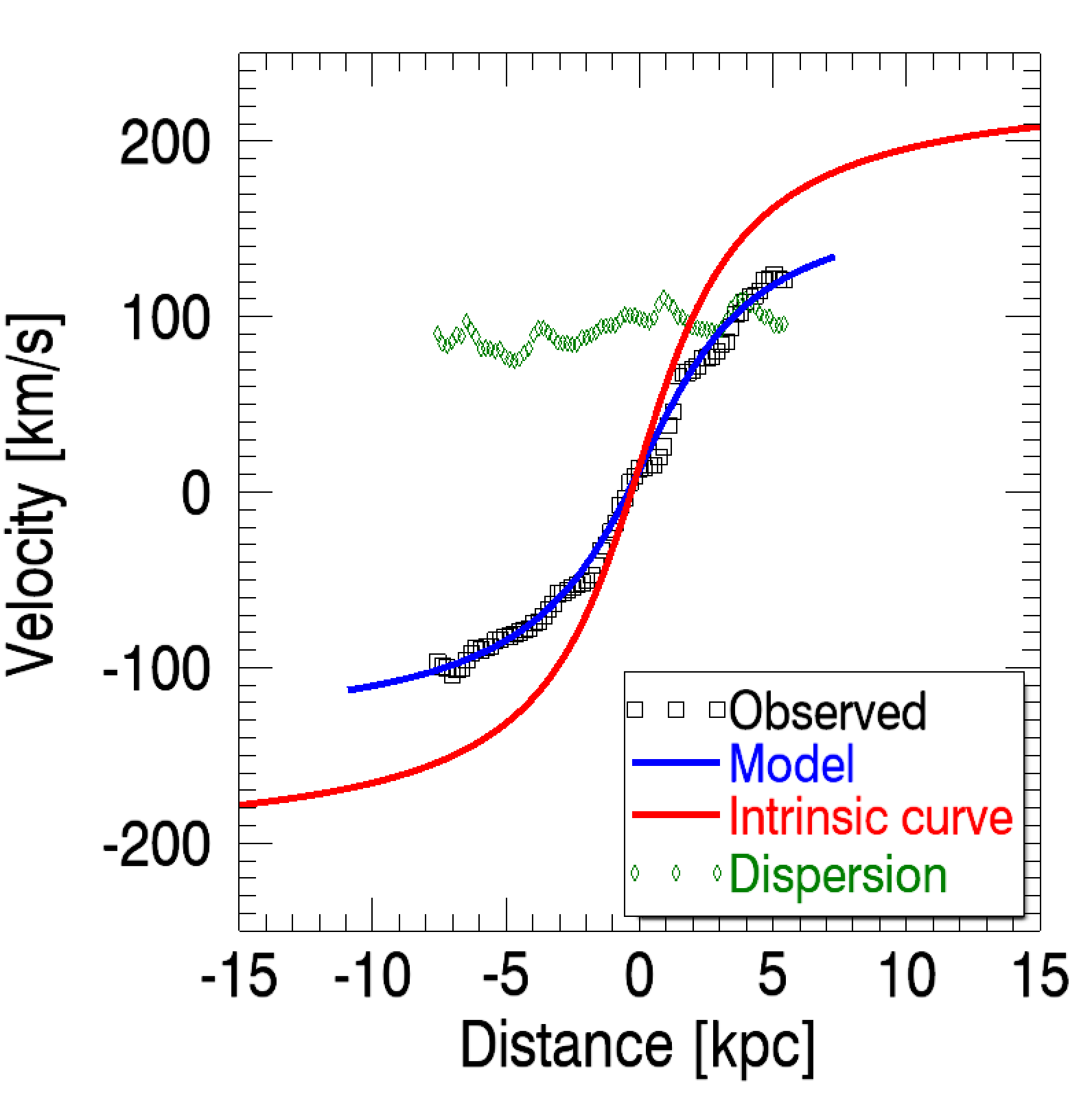}
      \caption{Rotation curve extracted on the major axis of the velocity maps. The black squares and the blue line represent the curve extracted from the observed velocity map and velocity map from the model, respectively. The red line shows the intrinsic rotation curve from the model, corrected for inclination. The green diamonds represent the velocity dispersion profile.
              }
         \label{rot_curve}
   \end{figure}

\section{Physical properties}
\label{Sect5}

\subsection{Star formation rate and attenuation comparisons} 
\label{Sect51}

We now analyze the H$\alpha$ derived star formation rate and the different measures of the attenuation
we obtain from the integrated spectrum of A68-HLS115. 
The main integrated properties are summarized in Table \ref{int}.
The redshift determined with the H$\alpha$ emission line of $z=1.58582\pm0.00005 $ is in good agreement with the redshift found with the CO emission (see Table \ref{CO}). 

In A68-HLS115, we are able to measure both the SFR$_{H\alpha}$ corrected from the dust attenuation using the Balmer decrement (H$\alpha$/H$\beta$ ratio) and the SFR(IR+UV) derived from the UV and IR luminosities. We find that SFR$_{H\alpha}$ is $\sim1.7$ times larger than SFR(IR+UV) when carefully taking into account the differential amplification effects by measuring the H$\alpha$ flux in the source plane. 

Given the relatively large uncertainty on the extinction correction from the Balmer decrement, the SFR$_{H\alpha}$ agrees, however, within $\sim 1.5 \sigma$ with SFR(IR+UV).
Relatively few star-forming galaxies at $z \ga 1$ have joint measurements of the Balmer decrement, SFR$_{H\alpha}$, and the UV+IR coverage
to determine  SFR(IR+UV) \citep[see e.g.][]{Price2014,Shivaei2016,Puglisi2017}.
Studying 12 $z \sim 1.6$ starburst galaxies with SFR(IR) $\sim 200-400$ \msunyr, approximately eight times above the main sequence, 
\citet{Puglisi2017} find that the extinction-corrected SFR$_{H\alpha}$ traces only a small fraction of the total SFR, 
comparable to cases of local ultraluminous infrared galaxies (ULIRGs).
Examining somewhat less extreme galaxies at $z \sim 2$, \citet{Shivaei2016}  find good agreement between 
Balmer-decrement-corrected SFR$_{H\alpha}$ and the total SFR(IR+UV). 
In our case,  A68-HLS115 is a galaxy selected for its Herschel detection \citep{Sklias2014}, with properties similar to those of 
\citet{Shivaei2016}. Although we  find that  SFR$_{H\alpha} >$  SFR(IR+UV), this is not incompatible (within the uncertainties)
with the results of \citet{Shivaei2016}, and the discrepancy is negligible if H$\alpha$ is extinction-corrected with the Galactic law from \citet{Cardelli1989}, as pointed out by \citet{Reddy2015}.

The measure of the Balmer-decrement and the IR and UV luminosities of our source allow us to determine the
attenuation of the ionized gas and the stellar continuum, as has been done extensively for nearby galaxies and 
also for different samples of galaxies at redshifts up to $z \sim 2$ in different ways
\citep[e.g.][]{Reddy2010,Yoshikawa2010,Kashino2013,Wuyts2013,Price2014,Reddy2015,De-Barros2016}.
We find a high attenuation of the Balmer lines, with $A_V=2.90\pm0.58$, whereas the ``energy balance"
between the IR and the UV, that is, \ \lir/\luv, yields a lower attenuation $A_V=1.58^{+0.30}_{-0.10}$ (cf.\ Tables \ref{CO} and \ref{int}).
The ratio of the $A_V$ values derived in this way is $\sim 0.5$,
comparable to that of classical studies of low-redshift galaxies \citep[cf.][]{Calzetti2000}.
Adopting the Galactic extinction law \citep[][]{Cardelli1989},
we obtain a somewhat lower $A_V=2.26 \pm 0.45$, marginally higher that
the value from the "energy balance".
Such differences between the nebular and stellar attenuation have also been found at $z \sim 1-2$, for example by \citet{Yoshikawa2010} and \citet{Price2014}.
\citet{Reddy2015} and \citet{De-Barros2016} argue that this difference increases with increasing SFR of the galaxies,
thus reconciling earlier studies with apparently discrepant conclusions on the nebular and stellar color excess.
The exact physical explanation for these observed differences between stellar and nebular attenuation and their
dependence on galaxy properties (e.g.,\ on SFR, sSFR, age and others) are still debated \citep[see e.g.][for conflicting views]{Price2014,Reddy2015}.

Finally we note that the attenuation of A68-HLS115 is fairly high, with an $A_V \sim 1.6-3$.
The observed Balmer decrement indeed corresponds to an attenuation of H$\alpha$ by a factor of approximately ten.
Although high, this attenuation is quite consistent with expectations from the average relation between $A_V$ and stellar mass,
which is found by numerous studies for galaxies at least out to $z \sim 3$ \citep[cf.][]{Dominguez2013,Price2014,Alvarez-Marquez2016}.
It should also be recalled that our source has been selected from the IR, which should favor dusty galaxies.

\subsection{Metallicity and metallicity gradient}

We can use two metallicity indicators in our galaxy: N2 using [NII]/H$\alpha$ and O3N2 using [NII]/H$\alpha$ and [OIII]/H$\beta$. They both lead to a consistent oxygen abundance of $8.58\pm0.07$ and $8.52\pm0.10$ (see Table \ref{int}), respectively, when using the indicators from \citet{Pettini2004}.

Figure \ref{met_gradient} (top panel) shows the oxygen abundance gradient as a function of radius obtained using the N2 indicator. It has been derived at three different radii by fitting elliptical annuli on the velocity map in the source plane with the method of \citet{Krajnovic2006}. By averaging metallicity values of all the spaxels in the annuli, we obtain a slightly negative metallicity gradient (higher metallicity in the core than in outer regions) of $-0.057\pm0.070 $ dex in total (or $-0.014\pm0.016$ dex  kpc$^{-1}$) although with a low significance. This is consistent with observations at high redshift which show generally flat or negative gradients \citep[e.g.][]{Jones2013, Stott2014}. 
However, it is now known that the spatial resolution and annular binning can have an impact on the obtained metallicity gradient \citep[][]{Yuan2013, Mast2014}. Indeed, since our observations are seeing-limited and considering that the FWHM of the PSF varies between 1 and 5 kpc depending on the orientation, 
the observed metallicity gradient could be flattened by these effects.
Using only three annuli to determine the gradient could also cause a flatter gradient.

Moreover, our metallicity gradient is similar to gradients found in galaxies with the same stellar mass and sSFR as shown in Fig. \ref{met_gradient} (middle and bottom panels). We include data from \citet{Swinbank2012}, \citet{Stott2014}, and \citet{Wuyts2016}, and the relation obtained by \citet{Stott2014} (in orange) for comparison. No correlation is seen with the stellar mass. 
However, \citet{Wang2017} observe a tentative anti-correlation between stellar mass and metallicity gradient, coherent with a scenario where more massive galaxies are more evolved. 
\citet{Stott2014} find a trend with the sSFR while combining their data with those of \citet{Rupke2010b}, \citet{Queyrel2012}, and \citet{Swinbank2012}, which could imply that the sSFR is driven by the amount of gas flowing towards the center. Using simulations, \citet{Sillero2017} find in a recent study that a correlation between the metallicity gradient and sSFR is seen in certain conditions when strong gas inflows are produced by interactions or instabilities. However, when adding the data from \citet{Wuyts2016}, this trend becomes less obvious, which could mean that the gas flowing to the center might not be the only physical process that influences the sSFR.

\subsection{Kinematics}

Figure \ref{kin_source} (left panel) presents the velocity map in the source plane centered on the redshift obtained from the H$\alpha$ emission line of the integrated spectrum. The map shows an obvious velocity gradient. To model the kinematics, we use a markov-chain Monte Carlo (MCMC) method to fit the observed velocity map with the PSF convolved model. We adopt the arctangent function for the velocity profile \citep{Courteau1997}, which has been used in many studies (e.g., \citeauthor{Jones2010} \citeyear{Jones2010}):

\begin{ceqn}
\begin{align}
\label{eq_arctangent}
\upsilon(r) = \upsilon_{rot} \frac{2}{\pi} \mathrm{arctan}\frac{r}{r_t},
\end{align}
\end{ceqn}
where $r$ is the radius, $r_t$ is the turnover radius, and $\upsilon_{rot}$ is the maximum rotation velocity. All the parameters are free to vary when fitting the model. The fit is performed in the source plane. The model velocity map obtained is shown in Fig. \ref{kin_source} (middle panel) and the kinematic properties obtained are presented in Table \ref{int}. 

The rotation curve determined from the reconstructed H$\alpha$ velocity map in the source plane and from the kinematic model is presented in Fig. \ref{rot_curve}. The curve in red indicates the intrinsic rotation curve, corrected for the inclination. The rotation curve reveals a rotating disk typical of star-forming galaxies on the MS (e.g., \citeauthor{ForsterSchreiber2009} \citeyear{ForsterSchreiber2009}; \citeauthor{Wisnioski2015} \citeyear{Wisnioski2015}).

The observed velocity dispersion map is shown in Fig. \ref{kin_source} (right panel).
The intrinsic velocity dispersion is measured in spaxels on the major axis in the outer region of the disk to avoid beam smearing as much as possible and is also corrected for the instrumental broadening following Eq. \ref{eq_sigma}. From \citet{Burkert2016} and \citet{Johnson2017}, we know that even if we take the measurement in the outer region, there is still an increase of the dispersion value due to the beam smearing, which is of $5$-$15\%$ if we take into account the stellar mass and inclination of our galaxy. If we add this extra-correction for the beam smearing, we obtain an intrinsic velocity dispersion of $\sigma_0 \sim 80\pm10$ km s$^{-1}$ (Table \ref{int}). This value is in agreement with an evolution of the intrinsic velocity dispersion with redshift, where galaxies at high redshift show a higher intrinsic velocity dispersion compared to the local galaxies
\citep[e.g.][]{Wisnioski2015,Turner2017, Girard2018}. This value is nevertheless higher compared to samples at $z=1$ and $z=2$ of \citet{Wisnioski2015}, which show a mean dispersion of 25 and 50 km s$^{-1}$, respectively. This suggests that the high turbulence observed in our galaxy could be due to the high gas fraction of $f_{gas}=0.75\pm0.15$ of this source.
Moreover, the observed velocity dispersion map does not show any evidence of a peak in the kinematic center. An increase of the velocity dispersion in the center can be an indication of a rotating disk according to \citet{Rodrigues2017}. However, this increase is lower for galaxies with high velocity dispersion ($>70$~km s$^{-1}$) and could explain why we do not see this effect here \citep{Johnson2017}.

We obtain a ratio $\upsilon_{rot}/\sigma_0= 2.73 $, meaning that the galaxy is dominated by rotation if we use the typical criterion of $\upsilon_{rot}/\sigma_0>1 $ \citep[e.g.][]{Wisnioski2015,Mason2017,Turner2017,Girard2018},  but this ratio also indicates that the disk is marginally stable. This value is typical at $ z\sim 1.6$ according to the stability ($\upsilon_{rot}/\sigma_0$) evolution relation with redshift obtained by \citet{Wisnioski2015}, and our galaxy is in good agreement with the trends they find where the stability ($\upsilon_{rot}/\sigma_0$) is correlated to sSFR and $f_{gas}$. Indeed, they get a lower stability ($\upsilon_{rot}/\sigma_0 \lesssim \ 3$) at high sSFR ($\rm \gtrsim 2 \, Gyr^{-1} $) and $f_{gas}$ ($ \gtrsim 0.5 $). This is fully in line with the $\upsilon_{rot}/\sigma_0$ , sSFR, and $f_{gas}$ measured in A68-HLS115. This suggests that the disk turbulence and instability in this galaxy are mostly regulated by incoming gas (available gas reservoir for star formation).

Moreover, the Toomre stability criterion \citep{Toomre1964} can be expressed as

\begin{ceqn}
\begin{align}
\label{Eqtoomre}
Q_{crit}=\sqrt{2}\frac{\sigma_0}{\upsilon_{rot} \, f_{gas}}.
\end{align}
\end{ceqn}
We obtain for our galaxy a direct measurement of $Q_{crit}=0.70$. A value of $Q_{crit}=1$ is expected for a {\it thin} quasi-stable gas disk 
and $Q_{crit}<1$ is an indication of an unstable system that can fragment into clumps \citep[e.g.][]{Jones2010}. 
However, the critical value of the Toomre parameter for a {\it thick} gas disk is $Q_{crit}=0.67$ \citep{Kim2007}, which could be the case of A68-HLS115 since several studies point out that disk galaxies at $z\sim1-2$ are thick \citep[e.g.][]{Genzel2011}. \citet{White2017} also obtain a linear correlation between $\sigma_0/\upsilon_{rot}$ and $f_{gas}$ from Eq. \ref{Eqtoomre}. From this relation and assuming hydrostatic equilibrium and that the pressure is caused only by the turbulent motions of the gas, they find that a higher gas fraction leads to thicker disks in marginally stable disks.

   
      \begin{figure}[tb]
   \centering
      \subfloat{\includegraphics[scale=0.35]{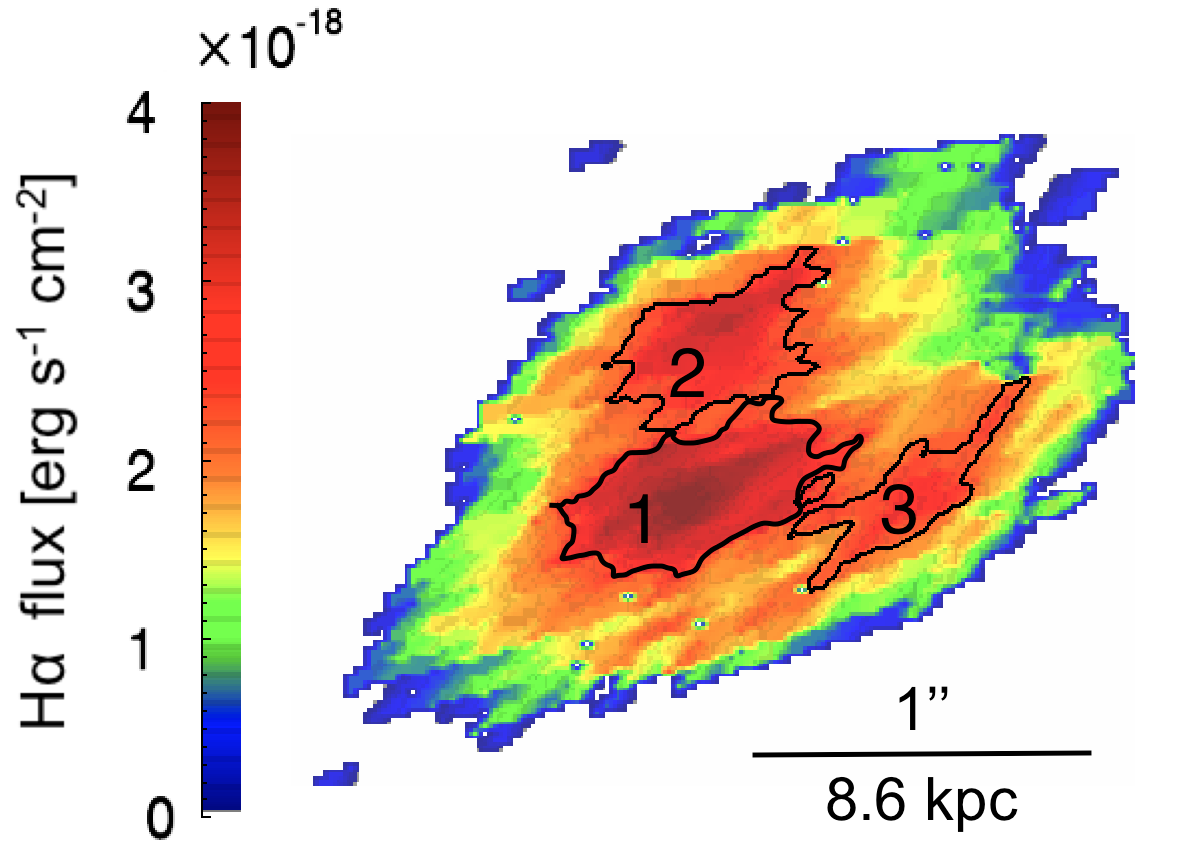}}\\
    \subfloat{\includegraphics[scale=0.35]{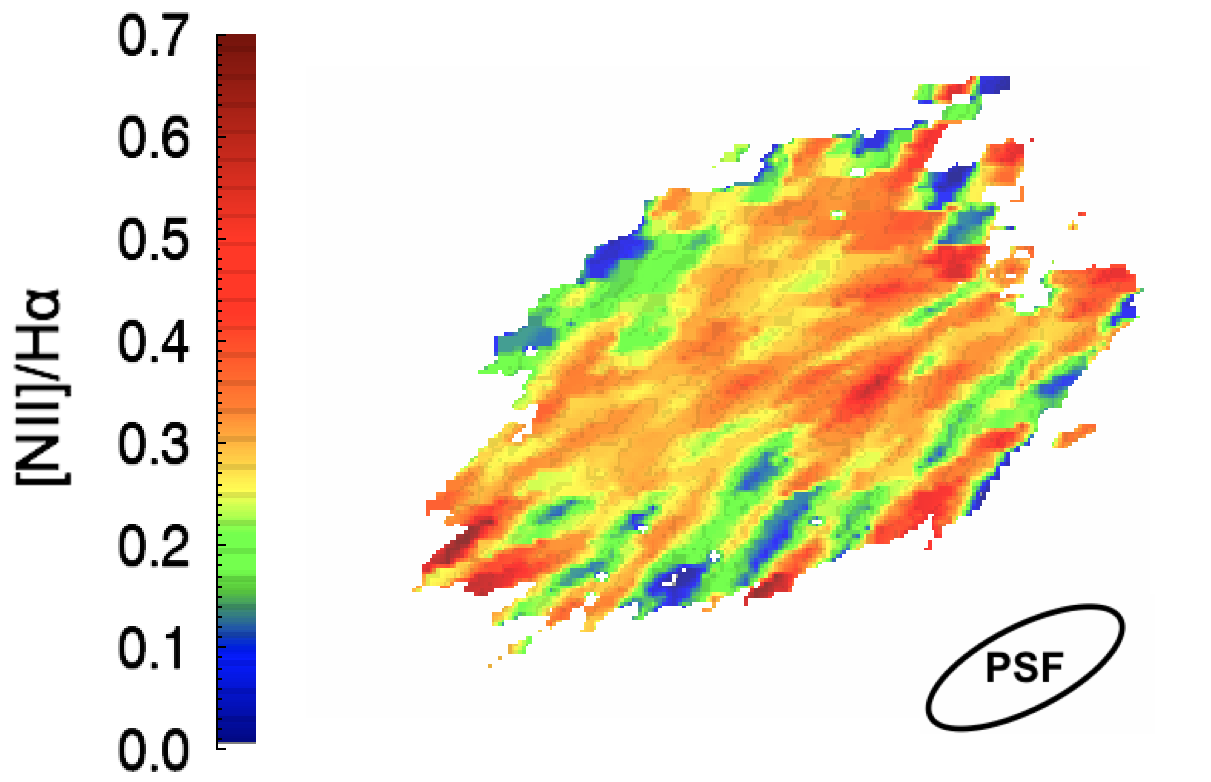}}\\
    \subfloat{\includegraphics[scale=0.46]{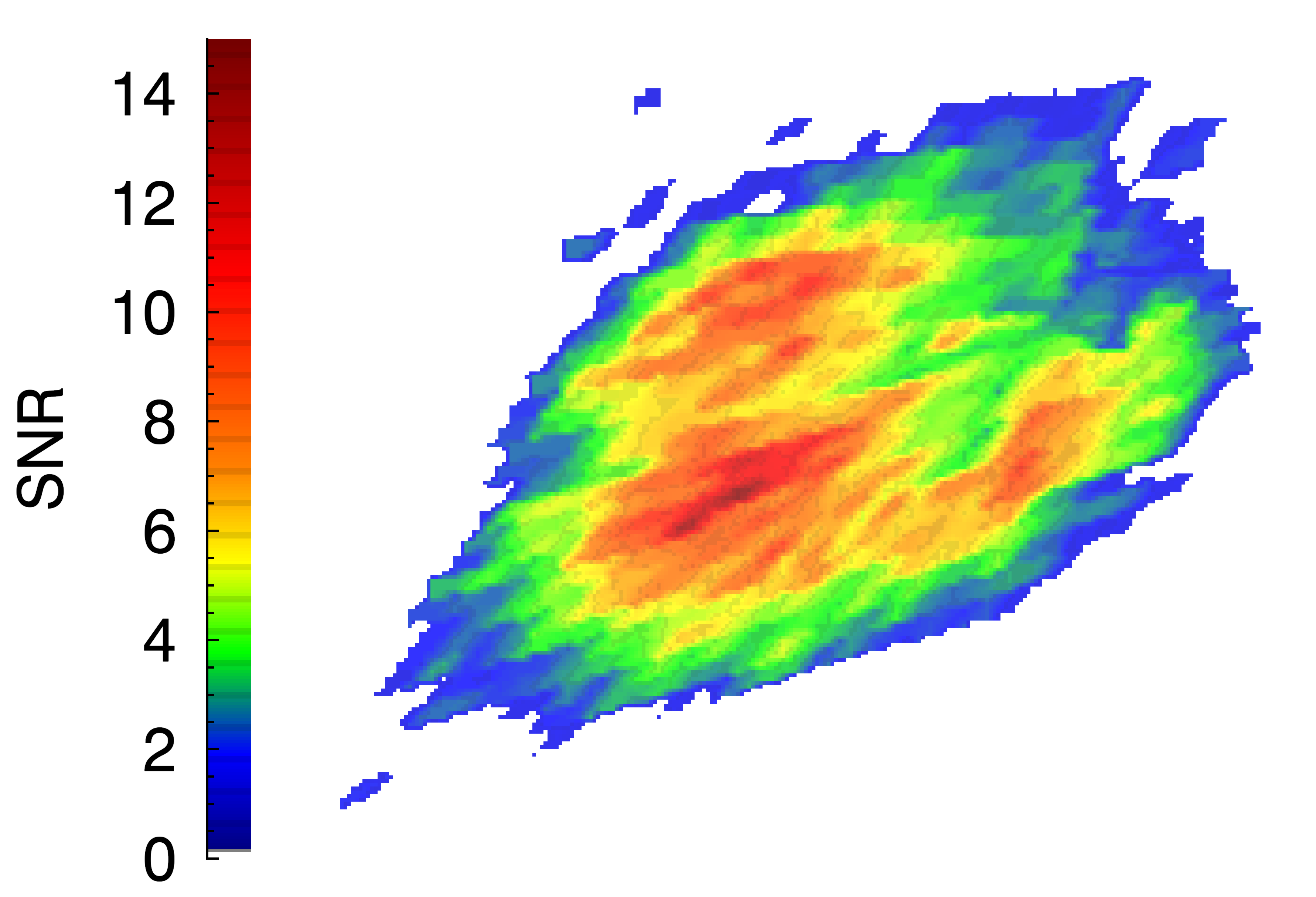}}
      \caption{
     H$\alpha$ flux per spaxel (top panel),  [NII]/H$\alpha$ line ratio (middle panel), and S/N (bottom panel) in the source plane. The black contours represent the H$\alpha$ clumps.
              }
         \label{flux_ha}
   \end{figure}

   
      \begin{figure*}[htb]
      \centering
   \includegraphics[scale=1]{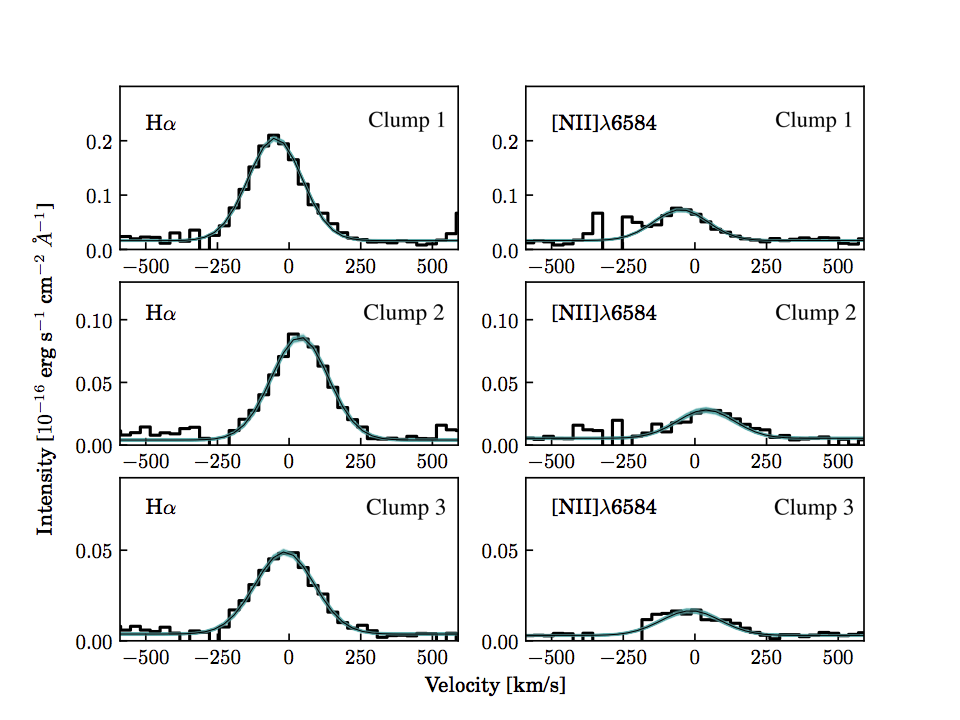}
    \caption{ H$\alpha$ and [NII] emission line intensities as a function of the velocity for the three clumps. The zero velocity corresponds to the redshift obtained with the H$\alpha$ emission line from the integrated spectrum. The 1000 realizations obtained with the Monte Carlo simulation are shown in blue and the final result is shown in black.}
         \label{clump_spectra}
   \end{figure*}

\subsection{Properties of the clumps}

The H$\alpha$ and [NII]/H$\alpha$ maps of A68-HLS115 in the source plane are presented in Fig. \ref{flux_ha}. We are able to identify three clumps in the image plane of the H$\alpha$ map with CLUMPFIND \citep{Williams1994} using a selection criterion of $\geq 3\sigma$. Overlaid on the observed HST/F814W images with dashed white lines (see Fig. \ref{HST_image}), we see that these H$\alpha$ clumps are blends of multiple smaller star-forming clumps resolved in the HST image. 
The H$\alpha$ clumps are exclusively distributed over the eastern part of the galaxy, while no H$\alpha$ clump is detected in the western part yet holding one of the most prominent HST clumps.
When looking in the source plane after the reconstruction, we find that the H$\alpha$ clumps as well as the clumps observed in the HST images could trace a spiral structure. This scenario would be consistent with the observed velocity map which is typical of a rotating disk.


\begin{table}
\caption{Clump properties.}           
\label{table_clump}      
\centering                          
\begin{tabular}{l c c c}        
\hline\hline                 
Parameters & Clump 1 & Clump 2 & Clump 3 \\    
\hline                        
            \noalign{\smallskip}
            SFR$_{H\alpha}$\tablefootmark{a}  $[\mathrm{M}_{\odot} \mathrm{yr}^{-1}] $  & $41\pm3$      & $24\pm2$       & $10\pm1$ \\
            Radius\tablefootmark{b}   [kpc] & 1.57      & 1.34       & $\leq$ 1.26     \\
            12+log(O/H)$_{N2}$         & $8.60\pm0.07$     & $8.58\pm0.07$        & $8.60\pm0.08$              \\
            12+log(O/H)$_{O3N2}$      & $8.55\pm0.11$       & -       & -             \\
         
             12+log(O/H)$_{O3N2,H\alpha}$\tablefootmark{c}      & $8.58\pm0.08$       & $8.52\pm0.08$       & $8.60\pm0.09$             \\
            $\rm \sigma \, [km \, s^{-1}]  $ & $92 \pm 2    $ & $97 \pm 3     $  & $101 \pm 3  $            \\
             $\rm \sigma_{med}\tablefootmark{d} \, [km \, s^{-1}]  $ & $90 \pm 2    $ & $95 \pm 1     $  & $101 \pm 2  $            \\
            \noalign{\smallskip}
\hline                                   
\end{tabular}
     \tablefoot{\\
       \tablefoottext{a}{ The SFR has been corrected for lensing and for the dust attenuation using the extinction measured from the integrated spectrum.}\\
     \tablefoottext{b}{The radius corresponds to $\sqrt{a\times b}$, where $a$ and $b$ are the semi-major and semi-minor axis. The typical error on the size of the clump is about $10\%$.}\\
       \tablefoottext{c}{Since the H$\beta$ emission line is only detected in the clump 1, the H$\beta$ flux has been derived from the H$\alpha$ flux using a H$\alpha$/H$\beta$ ratio of 2.86 and the $A_V$ measured from the integrated spectrum.}\\
     \tablefoottext{d}{$\sigma_{med}$ is the median dispersion value of the spaxels inside each clump. The errors correspond to 95\% confidence range.}
     }
\end{table}


We combine all the spaxels within each clump and obtain individual spectra. We fit the H$\alpha$ and [NII] emission lines in the same way as the integrated spectrum (see Sect. \ref{Sect4.2}) for each clump (see Fig. \ref{clump_spectra}). For the brightest clump (clump 1), it is also possible to fit the [OIII] and H$\beta$ emission lines. As a result, we obtain the total flux of each emission line for each clump, the SFR$_{H\alpha}$ assuming a uniform extinction correction derived from the integrated spectrum, the oxygen abundance using the N2 and O3N2 indicators from \citet{Pettini2004}, for which we use the H$\alpha$ flux to derive the H$\beta$ flux, and the velocity dispersion corrected for the instrumental broadening. Since the velocity dispersions can be affected by beam smearing, we also measure the median dispersion value per clump using the velocity dispersion map. Indeed, this effect is less important for the measurement of individual spaxels. The main properties of the clumps are summarized in Table \ref{table_clump}.

We get a very similar metallicity for each clump using the N2 indicator, while using the O3N2 we get a value slightly lower for clump 2, but the difference is not significant. 
We also get a high velocity dispersion for our three clumps, with a difference of $\rm \sim 11 \, km \,s^{-1}$ between clump 1 and 3.
Overall, the clumps show a similar metallicity, a similar velocity dispersion as in the rest of the disk (see Table \ref{int}), and seem to be embedded in the rotating disk.

\citet{Livermore2015} show an evolution of the clump surface brightness with higher values at high redshift compared to local HII regions. The three clumps follow this trend with an SFR$_{H\alpha}$ about $100$ times higher than what we find in the local Universe for clumps with similar size according to the relation obtained by \citet{Livermore2015} at $z=0$ using the SINGS survey  \citep{Kennicutt2003}. 
We also find that the three clumps contribute individually to $5-20\%$ on the SFR$_{H\alpha}$ of the whole galaxy. Therefore, the majority ($\sim63\%$) of the observed $H\alpha$ flux is found outside these regions. 
Similarly, \citet{Guo2012} obtain SFRs of $\sim10\%$ and $\sim50\%$ for individual clumps and total contribution, respectively, for star-forming galaxies at $z\sim2$ with SFR measured through SED fitting.

For clump 1 we can measure the Balmer decrement (H$\alpha$/H$\beta$=8) , which turns out to be higher than that measured from the integrated spectrum. The attenuation of clump 1 is therefore quite high, $A_V=3.6$, showing that at least this clump does not correspond to a region which is visible due to inhomogeneities in the dust distribution, which could mimic stellar and ionized gas clumps, as suggested by \citet{Buck2017}.
We therefore conclude that this ``clump" (region) does most likely indeed correspond to some physical concentration of ionized gas (HII region(s)), which is primarily powered by one or several young stellar clusters, and that such physical entities truly form in high-redshift galaxies.

\section{Conclusions}
\label{Sect6}

We have presented new observations of A68-HLS115 at $z=1.5858$, a star-forming galaxy located behind the galaxy cluster Abell 68, but strongly lensed by a cluster galaxy member, with SINFONI, a NIR IFU at the VLT. We detect H$\alpha$, H$\beta$, [NII], and [OIII] emission lines. Combined with images covering the B band to the FIR and CO(2-1) observations, this makes this galaxy one of the only sources for which such multi-band observations are available and for which it is possible to study the properties of resolved clumps and to perform a detailed analysis of the integrated properties, kinematics, and metallicity. This source is also one of the most gas-rich galaxies ($f_{gas}=75\%$) known at $z>1$.

A68-HLS115 is a dusty and gas-rich galaxy with a stellar mass $\mstar \sim 3 \times 10^{10}$ \msun\ and a star formation rate SFR~$\sim80 -120$~\msunyr, as measured from the UV, IR, and from SED fits (Table \ref{CO}). Comparing the attenuation from Balmer decrement with that derived from the UV and IR luminosity, we find a higher attenuation for the nebular lines compared to the stellar continuum, as found for several other galaxies at high redshift \citep[e.g.][]{Yoshikawa2010,Shivaei2016}.

The galaxy shows a velocity map typical of rotating galaxies and we obtain a high intrinsic velocity dispersion of $\rm 80\pm10 \, km \,s^{-1} $. The stability ratio of  $\upsilon_{rot}/\sigma_0=2.73$ is characteristic of galaxies at high redshift with a high sSFR and gas fraction and suggests that the disk is marginally stable. We obtain a direct measurement of the Toomre stability criterion of $Q_{crit}=0.70$, which could suggest the presence of a thick gas disk. We also find a slightly negative metallicity gradient.

We are able to identify three clumps in the H$\alpha$ map that show  similar metallicity and  velocity dispersion to one another, but also to the host galaxy, and that seem embedded in the rotating disk. We obtain star formation rate densities approximately $100$ times higher than what is found for HII regions in the local Universe. Finally, the clumps in our galaxy represent $\sim40\%$ of the SFR$_{H\alpha}$ of the whole galaxy.

For one of the H$\alpha$ clumps, we can measure the extinction from the Balmer decrement, finding an extinction which is higher than the average over the entire galaxy. This shows that at least this clump is not just an appearance caused by inhomogeneities in the dust distribution of the interstellar medium, as suggested by \cite{Buck2017} for the stellar or ionized gas clumps frequently observed at high redshift.

\begin{acknowledgements}
This work was supported by the Swiss National Science Foundation. MG is grateful to the Fonds de recherche du Qu\'ebec - Nature et Technologies (FRQNT) for financial support. 
     
\end{acknowledgements}

\bibliographystyle{aa} 
\bibliography{references} 

\end{document}